\documentclass[graybox]{svmult}

\usepackage{mathptmx}       
\usepackage{helvet}         
\usepackage{courier}        
\usepackage{type1cm}        
\usepackage{makeidx}         
\usepackage{graphicx}        
\usepackage{multicol}        
\usepackage[bottom]{footmisc}

\makeindex             

\begin{document}

\title*{Detection of chaos in RR Lyrae models}
\author{E. Plachy, Z. Koll\'ath and L. Moln\'ar}
\institute{Emese Plachy \at Department of Astronomy, E\"otv\"os University, P\'azm\'any P\'eter s\'et\'any 1/a, H-1117 Budapest, Hungary, \email{eplachy@astro.elte.hu}
\and Zolt\'an Koll\'ath \at Konkoly Observatory, MTA CSFK, Konkoly Thege Mikl\'os \'ut 15-17, H-1121 Budapest, Hungary \email{kollath@konkoly.hu} \and L\'aszl\'o Moln\'ar \at Konkoly Observatory, MTA CSFK, Konkoly Thege Mikl\'os \'ut 15-17, H-1121 Budapest, Hungary \email{lmolnar@konkoly.hu}}
%
%
\maketitle

\abstract*{The period doubling phenomenon was recently discovered in RR Lyrae stars with  
the Kepler space telescope and has been theoretically explained by  
hydrodynamic calculations. However, peculiar solutions of the  
Florida-Budapest turbulent convective hydrodynamic code suggest that  
bifurcation cascade may evolve to chaos in these dynamical systems.
We show that chaotic behaviour may be recovered from the radius variations of  
the model using the global flow reconstruction method. The fractal (Lyapunov)  
dimension of the underlying dynamical attractor is calculated to be ~2.2.  
Compared to the radius, the luminosity variations proved to be less suitable  
for such investigations due to their complexity. That suggest that even the  
continuous Kepler data would require transformation before conducting a  
similar analysis.}


\section{Motivation}
\label{sec:1}
Dissipative dynamical systems with relatively small linear growth rates are usually not expected to show chaotic behaviour. However, chaos investigations of these systems are justified in the light of new discoveries. The period doubling phenomenon was recently observed in RR Lyrae stars with the \textit{Kepler} space telescope (\cite{Szabo}) and has been explained by hydrodynamic calculations (\cite{Kollath}). The period doubling state is usually not ``far'' from chaos. We analyse two peculiar model solutions of the Florida-Budapest turbulent convective hydrodynamic code that suggest that the bifurcation cascade may evolve to chaos in these systems.

\begin{figure}[]
\begin{center}
\includegraphics[scale=.5]{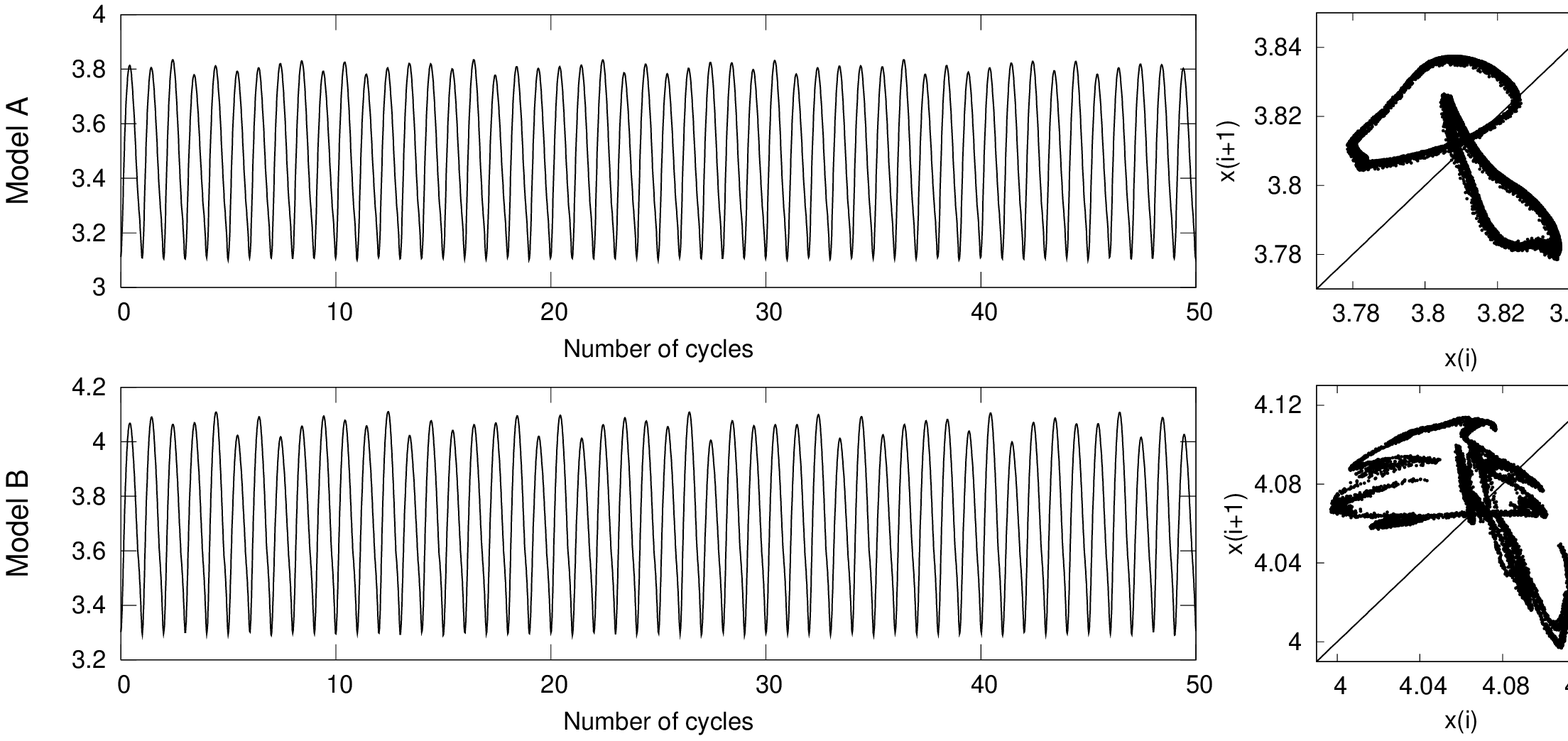}
\caption{Left panels: Radius variation of the two models. Right panels: Return maps for successive maxima.}
\label{fig:1} 
\end{center}      
\end{figure}

\section{Results}
We used the global flow reconstruction technique, a nonlinear analyser tool that is suitable to detect chaos and define quantitative information of the system  (\cite{Serre}). We have successfully reconstructed both models and determined the Lyapunov dimension to be $2.22\,\pm\,0.10$ in the case of Model A, and $2.17\,\pm\,0.08$ of Model B. These values are in agreement with the broad structure of the return maps on Fig.~\ref{fig:1}. 
Return maps display a more complex look compared to the usual quasi-one-dimensional tent or parabolic shape that chaotic systems have with Lyapunov dimension of $2+\epsilon$. We iterated the nonlinear models for $10^5$ cycles to rule out any transients but the return maps remained unaltered.

The kinetic energy changes less than a percent between pulsation cycles, in agreement with typical linear growth rates in RR Lyrae models.

Radius variations of RR Lyrae hydrodynamic models were suitable for the global flow reconstruction method and thus detection of chaotic behaviour. Luminosity variation was also studied in this manner, but the reconstruction was not successful. We believe that this is probably due to the more complex nature of the light curves. 

The observations of \textit{Kepler} RR Lyrae stars suggest some irregularity in the period doubling,
but a similar analysis can only be performed after a suitable transformation of the light variation.

\begin{acknowledgement}
The European Union and the European Social Fund have provided financial
support to the project under the grant agreement no. T\'AMOP-4.2.1/B-
09/1/KMR-2010-0003. This work has been supported by the Hungarian OTKA grant K83790.
\end{acknowledgement}

\end{document}